%% file: comalf.tex
\newdimen\sb \def\md{\sb=.01em \ifmmode $\rlap{.}$'$\kern -\sb$
                               \else \rlap{.}$'$\kern -\sb\fi}
\newdimen\sa \def\sd{\sa=.1em  \ifmmode $\rlap{.}$''$\kern -\sa$
                               \else \rlap{.}$''$\kern -\sa\fi}
\input stbasic.tex

\twelvepoint
\singlespace
\raggedbottom
\null\vskip.5in
\vskip.4in
\centerline{\bf  
The dwarf galaxy population of the Coma cluster to $M_R = -11$:
a detailed description}
\vskip.4in
\centerline{Neil Trentham }
\smallskip
\medskip
\centerline{Institute for Astronomy, University of Hawaii}
\vskip 4pt
\centerline{2680 Woodlawn Drive, Honolulu HI 96822, U.~S.~A.}
\vskip 4pt
\centerline{email : nat@newton.ifa.hawaii.edu}
\vskip.3in
\vskip.3in
\centerline{                   } 
\vskip.3in
\vskip.3in
\vfil
\eject
 
\centerline{\bf ABSTRACT }
\bigskip
\noindent
We present the luminosity function and measurements of the
scalelengths, colours, and radial distribution of dwarf
galaxies in the Coma cluster down to $R=24$.
Our survey area is 674 arcmin$^{2}$; this is the deepest
and most detailed survey covering such a large area.

Our measurements agree with those of most previous authors
at bright and intermediate magnitudes.  The new results are:

\vskip 3pt
\noindent
1) Galaxies in the Coma cluster have a luminosity function
$\phi (L) \propto L^{\alpha}$ that is steep
($\alpha \sim -1.7$) for $-15 < M_R < -11$, and
is shallower brighter than this.  The curvature in the
luminosity function 
at $M_R \sim -15$ is statistically significant.

\vskip 3pt
\noindent
2) The galaxies that contribute most strongly to the luminosity
function at $-14 < M_R < -12$ have colours and scalelengths 
that are consistent with those of local dwarf spheroidal galaxies
placed at the distance of Coma.

\vskip 3pt
\noindent
3) These galaxies with $-14 < M_R < -12$ have a colour
distribution that is very strongly peaked at $B-R = 1.3$.
This is suggestive of a substantial degree of homogeneity in
their star formation histories and metallicities. 

\vskip 3pt
\noindent
4) These galaxies with $-14 < M_R < -12$ also appear to be
more confined to the cluster core ($r \sim 200$ kpc)
than the brighter galaxies.  Alternatively, this observation
may be explained in part or whole by the presence of an
anomalously high number of background galaxies behind the
cluster core.  Velocity measurements of these galaxies would
distinguish between these two possibilities.


\vfil\eject

\noindent{\bf 1 INTRODUCTION}

\noindent
The Coma cluster (Abell 1656)
is the prototypical rich cluster of galaxies.  It is one
of two Abell richness $\geq 2$ 
clusters within 100 Mpc of us, and, of them, is the better 
studied (the other, Perseus = Abell 426,
lies in the Galactic plane).
In Table 1, we list its main properties.

The optical morphology of the Coma cluster has been well studied
(see, e.g.,~Rood \& Baum 1967). 
It has two central bright supergiant elliptical galaxies of 
approximately equal luminosity, NGC 4874
and NGC 4889; this may suggest that
Coma is the result of a merger between two smaller
subclusters.  NGC 4874 has a faint extended halo and has
been interpreted as 
a cD by Schombert (1988).
Coma has also been well studied in X-rays 
(e.g.,~Briel et al.~1992); it shows considerable
substructure $-$   
a dominant smooth
component
centered on NGC 4874 and a smaller component 41 arcminutes
to the southwest
centered on NGC 4839 (which Schombert also interprets as a
cD based on its extended halo).  
Coma is also notable for the absence of a cooling flow 
(Hughes et al.~1988);
Stewart et al.~(1984) suggest that a cooling flow there might have
been disrupted if Coma was the result of a merger of two smaller
subclusters.  

Because the galaxy density is so high, Coma has been
a popular place in which
to study the galaxy luminosity function (LF),
defined as the number density of galaxies per unit luminosity. 
Early attempts (e.g.,~Godwin \& Peach 1977, Lugger 1986)
concentrated on measuring the giant galaxy LF.  
Since then, we have come to realize that at lower luminosities
($M_B > -17$, or $M_R > -18$),
galaxies are almost exclusively members of
a separate population of galaxies,
the dwarfs (the dwarf spheroidals, or dSphs,
sometimes called dwarf ellipticals, and also the
dwarf irregulars, or dIrrs).
These are a different family of galaxies from giant ellipticals
(which are the most numerous kind of
giant galaxies in clusters; Dressler 1980) on the fundamental plane
parameter correlations (Kormendy 1985, 1987; Binggeli 1994).  
This suggests that different physical processes are at work in
producing their luminosities. 
Understanding the details of
these physical processes has driven much
recent theoretical activity.
The LF of dwarf galaxies 
depends mainly on the
efficiency with which gas is converted into stars in 
low mass systems (this efficiency depends on things like
the physics of wind ejection).  
Different models produce extremely
different results (compare Figure 14 of Babul \& Ferguson
1996 with Figure 2d of White \& Kauffmann 1994, for example). 
Therefore the measurement of the dwarf galaxy
LF offers a powerful probe of galaxy formation
theories.

The main observational difficulty in measuring the dwarf LF in
Coma has been to subtract the background 
galaxy number counts.
This is now a straightforward exercise, because 
large-format CCDs have permitted us to characterize well the
background galaxy number counts and their variance at
faint magnitudes (see, e.g.,~Driver 
et al.~1994b, Bernstein et al.~1995,
Trentham 1997a). 
The first work on this subject was done by Thompson \& Gregory
(1993), who determined the LF to $B\sim 20$ ($M_B \sim -15$) 
with photographic 
measurements.  Their results suggested a steepening of the LF
at their faint magnitudes: $\alpha \approx -1.4$. 
Subsequent measurements quickly followed.
Biviano et al.~(1995) made velocity measurments of a complete
sample of Coma galaxies down to $B=20$ and found a LF
that was rising at the faint-end. 
Karachentsev et al.~(1995) confirmed that there was an
excess of faint galaxies above the 
background and showed that these
excess galaxies had
systematically lower surface brightnesses
than the background galaxies.
Bernstein et al.~(1995) attempted to measure the Coma
LF to much fainter limits ($M_R \sim -9.5$) in
a smaller region of the cluster but
suffered from globular cluster contamination in the
halo of NGC 4874 at faint
magnitudes and could not constrain the galaxy LF strongly 
fainter than $M_R = -14$ in this region
(see their Figure 5).
More recently, Secker \& Harris (1996, hereafter SH96)
made a deep survey over a much larger area
(700 arcmin$^{2}$) and
found a LF that is rising to $M_R = -14$ and flattens fainter
than this.  Their best-fit LF has $\alpha = -1.4$. 
However, their faintest points depend on a completeness
correction; our simulations 
generally show these corrections to be
unreliable (see Trentham 1997a for a discussion of
correction methods and their pitfalls)
and very sensitive to details like the sizes of
the galaxies that are being corrected for.  
We need deeper imaging
of the fields that they studied to get results that are
independent of these corrections and to test
how valid these worries are. 
Lobo et al.~(1996) argue that $\alpha = -1.81 \pm 0.03$
for $-19 < M_V < -15$; it is unclear why a LF this
steep was not seen in the other surveys.

In this work, we extend the measurement of the LF 
1.5 magnitudes fainter than
SH96 in the $R$-band.  
We cover a similar area.
We also measure the $B-R$ colours and scalelengths of the
faintest dwarfs and investigate how
these properties vary with radial position in the
cluster.  We also attempt to compare them with nearby
dwarf galaxies. 

The results of Kormendy (1987) suggest that local dSph and dIrr
galaxies have exponential light profiles and $B$-band
absolute magnitudes $M_B$ and scalelengths $h$
that are related by the expression
$$M_B = -9.8 \log_{10}{h} - 16.6,$$ 
The central surface-brightnesses are 
$${\mu_B}_0 = M_B + 2.5 \log_{10}{2 \pi h^2}.$$
In the present work, our measurements are deep enough to detect all 
galaxies in the Coma cluster 
obeying this relation with $M_B < -11$ and $M_R < -12$,
assuming $B-R > 0.5$ (typically dIrrs have $B-R \approx 1$ and
dSphs have $B-R \approx 1.5$; see Trentham 1997b and
references therein).  The only such galaxies that we do not detect
are those that fall within the isophote of a much brighter galaxy 
(we correct for this).  In some of our survey area, we are able to
detect galaxies with absolute magnitudes 0.5 mag fainter than the limits
above because the image quality is better or because the sky brightness
is lower.   
We use a technique to measure total magnitudes that takes
into account the surface brightnesses of the galaxies, which
allows us (i) to get realistic estimates of the
uncertainties in total magnitudes (ii) to see clearly where
completeness worries might be important.  This method,
along with our observing details, is presented in Section 2.
In Section 3 we present and discuss the results.

Throughout this work we assume
that $H_0 = 75$ km s$^{-1}$ Mpc$^{-1}$. 

\vskip 10pt

\noindent{\bf 2 OBSERVATIONS AND PHOTOMETRY} 

\vskip 5pt

\noindent{\bf 2.1 Data collection and reduction}

\noindent
All observations presented here 
were taken
at the f/10 Cassegrain focus of the University of Hawaii 2.2 m telescope
on Mauna Kea during the nights of March 7$-9$ 1994, April 8$-$10 1994,
May 7$-$8 1994, and May 1$-$3 1995.
The detector was a 
Tektronix 2048 $\times$ 2048 CCD 
(scale 0\sd22 pixel$^{-1}$;
field of view
7\md5 $\times$ 7\md5).
The CCD was thinned and backside-illuminated and it had a high
quantum efficiency at short wavelengths, allowing deep imaging. 
We imaged a total of 654 square arcminutes of the cluster core
(see Figure 1) for 35 minutes in $B$ and 25 minutes in $R$ using
Mould filters, and a 56 square arcminute region centered on
NGC 4839 for 45 minutes in $B$ and $R$.
The imaging of the core region was done as a 4$\times$4 mosaic of  
individual fields.  
Each image was constructed from five shorter
exposures, dithered in order to reject bad pixels and to construct
sky flats.
A substantial dither (30 arcseconds) was needed so that 
satisfactory 
median sky flats could be constructed for these fields, 
which contained
several large galaxies.
Hence the total survey area is somewhat smaller than
the total area of sixteen, 7\md5 $\times$ 7\md5 frames. 
The individual exposures were 
bias-subtracted and then flatfielded using a flatfield that
was constructed using both 
median sky and twilight flats.  
This produced images flat to better than one percent.
The CCD read noise was sufficiently low (18 e$^{-}$, gain
3.5 e$^{-}$/ADU) that all the images were dominated by
photon noise from the 
sky.  The seeing varied from 0\sd7 to 1\sd1 
FWHM.

Instrumental magnitudes were computed from observations of 
standard stars ($\sim 30$ per night).
The photometry was converted to
the Johnson (UBV) $-$ Cousins (RI)
magnitude system of Landolt (1992).  This conversion is accurate to
about 2\%.  Because we need to do an accurate background
subtraction to determine the LF, and because the background
number counts are a steep function of magnitude, an accurate 
zero-point is important. 
We need an uncertainty in
the relative zero-point between our cluster and background
fields $\Delta m_{\rm zpt} < 0.035$ mag in $R$ 
to ensure that the field-to-field variance remains the dominant source of
error.  Although our zero-point is only 2\% accurate relative to an
absolute scale, the uncertainty in the relative zero-points between
cluster and field will be much less than this (the same standards were
used to calibrate both datasets), and the inequality above is
easily satisfied.

Conditions were photometric when most of the images were taken; in the
few cases when there was thin cirrus overhead, the images were
calibrated using shorter exposures taken under photometric conditions.  

\vskip 5pt

\noindent{\bf 2.2 Photometry Techniques}

\vskip 5pt

\noindent{\bf 2.2.1 Computation of Total Magnitudes and
the Luminosity Function} 

\noindent 
The surface-brightness$-$dependent technique of making isophotal
corrections that we use to measure our galaxy magnitudes
is described in detail elsewhere (Trentham 1997a,
hereafter T97).  A detailed
evaluation of the technique as applied to real data is
presented there also. 
The main steps are:
\vskip 1pt
\noindent 1) We measure the rms sky noise $\sigma_{rms}$ and
the seeing FWHM $b_{\rm FWHM}$ for each image.
\vskip 1pt
\noindent 2) We then simulate galaxies of various
apparent magnitudes and exponential scalelengths.
These simulated galaxies are then
convolved with a Gaussian seeing function of width 
$b_{\rm FWHM}$, and  
Poisson noise of rms magnitude $\sigma_{rms}$ is added. 
In these simulations we consider all regions of the 
magnitude--scalelength plane 
excluding objects that are smaller than point sources and objects
that are too big and/or too faint to be detected above the
sky noise.  Figure 2 of T97 shows a typical such region of
this plane.
\vskip 1pt
\noindent 3) We then run the FOCAS detection algorithm 
(Jarvis \& Tyson 1981; Valdes 1982, 1989)
to search for objects with fluxes
that are 3$\sigma_{\rm rms}$ above the sky.
For each object we measure the isophotal magnitude $m_I$ and its
first-moment light radius
$r_{1} = \int r
I(${\bf r}$) {\rm d}^{2}${\bf r}/$\int
I(${\bf r}$) {\rm d}^{2}${\bf r}, where {\bf r} is the radial
vector measured in the plane of the sky from the center of
the object and $I$({\bf r}) is the projected light distribution
of the object.
As the true magnitude $m$ of each object is known
in the simulation, we compute
the function $m(m_I, r_1)$, and its uncertainty  
$\sigma (m)[m_I, r_1]$.
The uncertainty comes from studying how intrinsically
identical galaxies are detected differently depending on
the local noise.   
We also determine the faintest magnitude $m_L$ at
which galaxies whose intrinsic magnitudes and scalelengths
are equal to those of local dwarf galaxies seen at the
distance of Coma are detected with 100\% completeness
in the absence of brighter galaxies (the magnitude
vs.~scalelength relation from Section 1 was used). 
This will be the faintest magnitude to which we determine
the LF in each image.  We make this cut because the
completeness is a very strong function of magnitude
fainter than $m_L$ (it typically drops from 100\% to 
0\% over a range of one magnitude). 
Therefore completeness corrections would be very unreliable.
\vskip 1pt
\noindent 4) We then run the same detection algorithm
on our data and make a catalog of all objects detected
at the 3$\sigma$ level (typically 26 mag arcsec$^{-2}$) 
above the sky.  For each
object, we measure $m_I$ 
and $r_{1}$. 
\vskip 1pt
\noindent 5)  
Multiple objects within a single detection isophote
are identified by searching for
multiple maxima and are split into individual objects using
the FOCAS splitting algorithm.  The algorithm is run several times,
so that cases where many objects were contained in a single
isophote initially are all individually recovered. 
The quantities $m_I$ and $r_1$ are
computed for each object at each stage of the splitting.
\vskip 1pt
\noindent 6)
Objects are then classified 
based on their morphology relative to that of
several reference PSF stars in the field
(see Valdes 1989 for the details of the
classification terminology).
\vskip 1pt
\noindent 7)
We then remove from the 
catalog (i) objects with $m (m_I, r_1) < m_L$;  
(ii) diffraction spikes of bright stars, ghost images, and chip defects
(these were identified from the FOCAS classification $-$ see 
T97 and
Valdes 1989 for details);
and
(iii) objects in the halos of a number of giant elliptical or S0
galaxies having a associated population 
of red unresolved faint objects that
are probably globular clusters.  Table 2 lists 
the relevant galaxies and excluded areas.  This is done so that
the globular cluster contamination in our faintest one or two
bins is small $-$ Bernstein et al.~1995 showed that globular
cluster contamination 
is probably very significant at $R > 24$ in the halo of NGC 4874.
We also removed  
(iv) spurious objects that were detected in the halos of bright stars
and galaxies where the noise is much higher than for the rest of the
image.
The last of these
corrections required us to look at all objects in our image that
were part of a larger object that was subsequently split into
more than three smaller objects in our original detection pass and
to make a judgment by eye as to whether the faint objects we see are
really galaxies or stars, or local noise peaks that have been
enhanced because of the higher sky background in the halo of 
a bigger galaxy.
This was the most time-consuming part of this project, but the effort
spent is
justified by the increased confidence that
all objects that remain in our final
catalog are real.
Also, at this stage, a number of low-surface-brightness galaxies
that had been recognized as a multiple object and split into many
small objects centred on local noise peaks
were reconstructed, and the values of $m_I$ and $r_1$ that
were computed
prior to splitting were adopted.
\vskip 1pt
\noindent 8)
After these corrections are made, we then have a FOCAS catalog
of objects classified as ``galaxies''
or ``stars''.  At faint magnitudes, these
classifications are unreliable because many galaxies
have apparent scalelengths
smaller than the seeing and so look like stars.   
We therefore correct for stellar contamination in our images by
assuming that the shape of 
the Galactic stellar luminosity function (SLF) is invariant for
$20 < m < 25$, adopting
the measured SLF shape of  
Jones et al.~(1991) and computing the number of faint stars as a function of
magnitude 
based on this SLF normalized by the numbers of stars
detected at the bright end 
where the classifications are 100\% reliable.
For Coma, which  lies well out of the Galactic plane, this
is a small effect ($<5$\%), and the uncertainties generated by this
method are negligible.
\vskip 1pt
\noindent 9)
From our measured $m_I$ and $r_1$ values, for each object in our catalog we
then compute $m$ and its uncertainty $\sigma (m)$,
corrected for stellar contamination as in 8).
We bin the results in half-magnitude intervals.
The number counts (in units of number per half-magnitude per square degree)
are then computed by dividing the number of 
galaxies in each bin by the survey
area.  This survey area includes a correction for crowding, the
process by which faint galaxies go undetected because they happen to
fall within the detection isophote of a much brighter object (see 
T97 for details of how this was done).
The correction varies from about 30\% (in the one or two frames
that contain a supergiant galaxy like NGC 4889) to about 5\% (in
most fields).
\vskip 1pt
\noindent 10)
The number counts were corrected for Galactic extinction using
the HI maps of Burstein \& Heiles (1982) and the colour conversions of 
Cardelli et al.~(1989).  The Galactic extinction
is $A_B = 0.05$ mag for the core fields
and $A_B = 0.03$ for the NGC 4839 field.
Extinction from dust in Coma and
the effects on the background galaxies of gravitational lensing
by the cluster dark matter are assumed to be negligible
(see Bernstein et al.~1995). 

The numbers of galaxies that we detected in each bin are presented
in Tables 3 and 4.
Different fields had different $m_L$ values because the seeing varied; 
this is why the numbers 
of fields in the last rows of the table are
smaller than in the other
rows.
Our resulting magnitude $-$ number count
plots are shown in Figure 2 for the Coma core field and Figure 3
for the NGC 4839 field.  The uncertainties are the
quadrature sum of 
uncertainties from counting statistics and uncertainties from the
isophotal corrections $\sigma (m)$.  The uncertainties from counting
statistics dominate at the bright end, and the uncertainties from 
isophotal corrections dominate at the faint end.
Also shown in Figures 2 and 3
are the mean background counts for random 
sky fields (T97);   
it is clear from Figure 2 that for the Coma core region,
the background contributes more to the total number counts
at fainter magnitudes.

The luminosity function is computed
by subtracting the background contribution from the number counts,
and the uncertainty is computed taking into account the
field-to-field variance in the background in addition to the errors
described in the last paragraph. 

The background counts are presented in T97; we use both the
mean number counts and the field-to-field variance of
the background computed there (corrected for differences
in field size using Poisson statistics) in this work.
The background fields studied in T97 comprise a total area of 132.4
arcmin$^2$ in $B$ and 188.1 arcmin$^2$ in $R$ imaged to 
$m = 25.5$ in both passbands, which is deeper than any of
the Coma data presented here.  The background data was treated in
exactly the same way as the data in this work, as outlined earlier
in this section.  The background fields, like Coma, are at high
Galactic latitude ($b > 30^{\circ}$) 
so that stellar contamination is small there too.  We adopt
the field-to-field variance measurements of Bernstein et al.~(1995)
in the $R$-band, as they have slightly better statistics than we do, and
derive the $B$-band variance from the $R$-band variance using the
measurments of Driver et al.~(1994a; see T97 for details).

In converting apparent magnitudes to absolute
magnitudes, we adopt a distance modulus of 34.83.
  
\vskip 5pt

\noindent{\bf 2.2.2 Computation of Colours}

\noindent
For each object described above, we compute the aperture magnitude
$m_a$ in an aperture of diameter 3\sd0.
For the faint objects whose colours we will be measuring, we
take the colour $B-R$ as $(m_a)_B - (m_a)_R$.
Isophotal magnitudes are not used for computing colours 
because the detection isophotes are not the same in
the different bands.
The 3\sd0 aperture 
is large enough that differential seeing between
our $B$ and $R$ images and noise-induced subarcsecond
offsets between the light centers of objects in the two images
are negligible.  

The above method fails to give accurate colours for bright galaxies if
colour gradients are large.  However, we will only
present colours for galaxies with $R > 20.5$.

\vskip 10pt

\noindent{\bf 3 RESULTS AND DISCUSSION}

\vskip 5pt

\noindent{\bf 3.1 Luminosity Functions}

\noindent
We present the LF for the Coma core field 
in Figures 4 ($R$-band) and 5 ($B$-band).
The same general form of the LF is apparent in all the figures: a
steep drop towards the bright end
at $M_B < -21$, a shallow
rise toward lower luminosities at 
intermediate magnitudes ($-20 < M_B < -16$), and a steep rise 
fainter than this.  
The statistics are good enough that the curvatures at both 
the bright and
faint ends of the LF are highly
significant $-$ neither a power law nor a Schechter (1976) 
function provides an acceptable fit to the data. 

The data presented in Figures 4 and 5 are also
presented in Tables 5 and 6, along with the
local LF slope, measured in four different ways.
Here $\alpha$ is defined as  
the logarithmic slope of the LF: 
$\phi (L) \propto L^{\alpha}$ so that 
$N(M) \propto - \phi(L) { {{\rm d}L}\over{{\rm d}M}} 
\propto 10^{-0.4 (\alpha+1) M}$.
We define $-0.4 (\alpha_{mn} (M) + 1)$ as the
slope at $M$ of the best fitting polynomial of order $m$ to
the $2n+1$ points in
Figures 4(a) or 5(a), computed at 0.5 magnitude intervals, 
centered on $M$.  The case $m=1$ is a power-law
fit.  The error in 
$\alpha$ is mostly systematic; 
a good estimate of the true $\alpha$ is the median
of the four values and a plausible estimate
of the uncertainty in the
true $\alpha$ is the range of the numbers. 

The error bars get larger at fainter magnitudes because the
background counts contribute more to the total
counts at a rate faster than the field-to-field
variance decreases with increasing magnitude.
It is intriguing that the faintest point in the $R$-band is so low.   
We may be seeing a hint of a true turnover in the LF.
An alternative explanation is that this is a normalization
effect (recall that only 6 of the 16 fields that made up the
mosaic are included in this point).
It is therefore instructive to consider the counts in these
6 fields in isolation.
For these 6 fields considered in isolation, in the
$M_R = -11.11$ point, following background subtraction,
we have $7_{-185}^{+190}$ galaxies, and in the
$M_R = -11.61$ point 
we have $341_{-169}^{+103}$ galaxies.
If the errors are Gaussian, the probability of there being 
more galaxies in the fainter point is than
only 9.4\% $-$ this is significant, but only at
the 1.7$\sigma$ level.
 
Similar LFs have been seen by us,
but only at the faint end, in Abell 262 and the
NGC 507 Group (T97), and only at the bright end 
in Abell 665 and Abell 963 (Trentham 1997b).
We can only see the faint-end of the LF in Abell 262 and the NGC 507
Group because these are poor clusters so that counting statistics
at the bright end are poor so that we cannot constrain the LF.  
We can only see the bright-end of
the LF in Abell 963 and Abell 665 because the clusters are distant
($z=0.2$) and we cannot image them deeply enough to reach $M_B > -14$.
In a separate paper (Trentham 1997c), we combine these results with 
others from the literature and show that all the LF shapes 
are consistent with each other to a high degree of precision and
that the cluster LF appears to be universal. 

The upturn in the LF occurs faintward of the 
magnitude where the LF in Virgo changes from being giant-dominated
to dwarf-dominated ($M_B \sim -17$ for $H_0 =
75$ km s$^{-1}$ Mpc$^{-1}$,
which corresponds to $M_R \sim -18.5$ for dSph galaxies;
Binggeli 1987). 
In both $B$ and $R$, the LF is in fact quite
shallow at this transition magnitude.  This suggests that (i) either
the transition occurs at different magnitudes in different environments,
or (ii) the giant galaxy LF is decreasing by an amount only just less
than the dwarf-galaxy LF is rising at the transition magnitude $-$ hence
a total galaxy LF that is gradually rising and almost
featureless.  In Virgo, the type-specific
LF has been measured (Sandage et al.~1985), and suggests that
(ii) is the explanation there.  If the cluster LF is universal, this
would suggest that this explanation is valid here, too.  
   
Neither Figure 4 nor Figure 5 shows a significant
difference in the LF between the inner
($r<200$ kpc) and outer ($r>200$ kpc)
regions of the cluster.  The same parent LF (e.g.,~the composite LF
of Trentham 1997c) can fit both LFs individually at a high level of
confidence 
if the normalization is adjusted. 
The distribution of the dwarfs in the cluster is described in 
detail in Section 3.4.
There we do see an apparent difference between the distributions of giants
and dwarfs in the cluster at a low level,
and we give possible explanations. 

In Figures 4 and 5 we also present the LF of extreme low-surface-brightness
(LSB) galaxies.  How such galaxies are defined is described in the
next section, but these very diffuse galaxies, while almost
certainly cluster members, do not contribute  
most of the excess galaxies seen at faint magnitudes.
At brighter magnitudes they contribute proportionately more
to the LF ($\sim$ 60\% at $M_R = -17$ compared to $\sim$ 20\%
at $M_R = -12$).  This suggests that cluster dwarfs are easier
to distinguish from background galaxies on morphological
grounds alone at progressively brighter magnitudes.

The LF of the NGC 4839 Group is poorly constrained because the
group is sufficiently 
diffuse that 
the ratio of members to background galaxies at a given apparent magnitude is
low.  The field-to-field variance of the background then causes the
error bars to be large and we cannot measure the LF.  

\vskip 5pt

\noindent{\bf 3.2 Scalelengths}

\noindent 
In Figure 6, we present the $m_I$ and $r_1$ values of all the
objects we detect in the core field.  
These diagrams are similar to projections of the fundamental 
plane, except that the radius parameter is seeing convolved and the
magnitude parameter 
is a measured isophotal one, not a computed total
one.  Lower surface-brightness galaxies have a larger $r_1$ for
a given $m_I$.
Stars are included $-$ these have $r_1$ close to the seeing,
and have the lowest $r_1$  
of any objects of a given $m_I$.  The thickness of the band of points
at bright magnitudes and low $r_1$ gives a measure of the
range of the seeing in our fields. 

The solid line represents 
galaxies having the scalelengths of typical local
dSphs as seen at the distance of Coma and observed under similar
conditions to our data.  The local dSph fundamental plane is
not a tight correlation, 
and the scatter around this line is substantial.
The figure suggests that the excess objects we find at
$R > 19$ and $B > 20$ (galaxies with $M_R > -15.83$ and
$M_B > -14.83$ if they are in the cluster) 
in Figures 4 and 5 have scale-lengths consistent with those
of dSphs. 

We cannot determine which points in Figure 6 refer to cluster
galaxies and which to background galaxies. 
Comparison between the solid and dotted lines suggests that 
for much of our magnitude range of interest, cluster dwarfs 
have scalelengths similar to background galaxies.
Therefore a detailed quantitative
analysis of 
exactly which region of the fundamental plane is occupied by
dwarfs is not possible with our data. 
A natural follow-up program would be a
velocity survey of the faint galaxies with $B > 20$.
Then we could identify which galaxies are cluster members and
see where they lie in the magnitude$-$radius projection 
of the fundamental
plane (higher resolution imaging data would be required too).

The figure also suggests that in Coma, lower surface-brightness
galaxies do not contribute proportionately more to the LF at
fainter magnitudes.  We also found this to be true in Abell 262
and the NGC 507 Group (T97).
We find a number of galaxies whose surface brightnesses are
lower than those of any background galaxies i.e.,~those above the
dashed line in the figure.  
In Figures 4 and 5 we present the LFs for these objects in
panel (d).
That these galaxies have a flat
luminosity function confirms the earlier statement that the
LSB galaxies do not contribute more to the LF at
fainter magnitudes.  This statement is, however, subject to the
important caveat that our detection efficiency of LSB galaxies
decreases at fainter magnitudes, so that such galaxies 
might exist and not be seen. 
However, we do not find proportionately
more galaxies of lower surface-brightnesses within the
surface-brightness range where we can detect with 100\%
completeness, on going to fainter magnitudes i.e.~there are not
many points near the points with the
largest $r_1$ at the faintest $m_I$ that we do observe.  
This of course does not
rule out the existence of  
a population of {\bf very} low surface-brightness
objects at $20 < m < 24$ ($-15 < M_R < -11$ in Coma), 
but such objects have not
turned up in substantial number in any known environment.  
The faintest giant galaxies known, with 
central surface-brightnesses
of $\mu_R \sim 26.5$ mag arcsec$^{-2}$
(e.g.,~GP1444, Davies et al.~1988), 
would have been marginally detected here if they
were in Coma. 
  
Although the faintest galaxies we detect have scale lengths consistent
with those we would expect of local dSphs placed in Coma, this is
not sufficient information for us to conclude that they {\bf are}
dSphs.  This is because the dSph and dIrr fundamental planes 
overlap (Kormendy 1987, Binggeli 1994).   
We need additional information, in particular colours, to distinguish
between these two types of galaxies.  This is discussed in the
next section.
\vskip 5pt

\noindent{\bf 3.3 Colours}

\noindent
Figure 7 presents the colour distribution of galaxies with
$20.5 < R < 22.5$, computed as described in Section 2.2.2.
We selected this magnitude range 
because we want to examine the properties of the
galaxies where the LF is steeply rising. 
We only consider galaxies with $R < 22.5$ so as to ensure
completeness except for galaxies with $B-R>2.3$, which is
much redder than the colour of 
old stellar populations at the distance of Coma. 

Figure 7(b) shows the colour distribution after the estimated
contribution from the background has been subtracted.
That the
histogram is approximately zero at red colours, 
where the cluster
contribution is presumably negligible, is encouraging that
the subtraction is reliable.
This figure suggests that the Coma dwarfs have a colour
distribution that is strongly peaked at $B-R = 1.3$.
This would suggest that they are dSphs (dIrr's
have colours bluer than this: $B-R < 1.1$). 

Local dSphs have a substantial range of colours 
(see Hodge 1989, Trentham 1997b).  The main 
parameters determining the colour (Hodge 1989, 1994) are 
the star formation history (galaxies that have current or
recent star formation are bluer) and the metallicity
(higher metallicity stellar systems are redder).  
A value of $B-R = 1.3$
is near the blue end of the colours seen for local
dSphs and suggests that the Coma dwarfs might have
undergone more recent star formation than their
redder counterparts in Virgo 
(Caldwell 1983) or Abell 262 (T97). 
That the distribution is so peaked suggests
that most of these dwarfs would have had similar
star-formation histories to each other (and probably
similar metallicites too).  If the Coma cluster
really is the result of a recent
merger of two smaller clusters,
as suggested by its two large central galaxies
and absence of a cooling flow (Stewart et al.~1984),
perhaps star formation was triggered at this time 
in most 
low-mass galaxies, which might have had somewhat
more gas than they do now, by the
rapidly changing gravitational fields
during the merger. 

Figures 7(c) and 7(d) suggest a radial colour
gradient that is much smaller than that measured
by Secker (1996, hereafter S96) for brighter dwarf
galaxies.
We find, in the units of S96, 
$ { {\Delta (B-R)}\over{\Delta \log_{10} r}} 
= -0.0284 \pm 0.0208$.
S96 found $ { {\Delta (B-R)}\over{\Delta \log_{10} r}}
= -0.08 \pm 0.02$ for Coma dSph's with $15.5 < R < 20$.
What these results suggest is that
colour gradients get smaller on going to fainter
magnitudes.  
This could be because star formation is turned off
more efficiently on lower mass systems as they
enter the cluster environment.
Alternatively, the difference between the S96 results
and ours could be due to contamination of his sample
by red low-luminosity 
normal elliptical galaxies (like M32)
near the cluster center.  Dressler (1980) showed that
ellipticals prefer high density environments.
Such galaxies do not exist or are extremely rare
as faint as $M_R = -14.3$ ($R = 20.5$ in Coma), so
our sample will not suffer from this problem. 
S96 also found that the mean colour of the
$15.5 < R < 20$ dwarfs at 1$^{\prime}$ from
the cluster center is $B-R = 1.46 \pm 0.02$, slightly
redder than the mean value we find for the
$20.5 < R < 22.5$ sample of $B-R = 1.34 \pm 0.01$
(adopting the same colour criterion as Secker for
inclusion in the computation: $1.05 < B-R < 1.6$;
note that there is no significant colour gradient
in our sample). 
This does not, however, help to distinguish between
the two possibilites above. 

For the NGC 4839 group, as in Section 3.1, no
useful conclusions can be made because the galaxy 
density relative to the background is too low
and counting statistics dominate following
subtraction.

The LSB galaxy colour distribution (Figure 7(f)) is
also peaked at $B-R = 1.3$.  
These too may be dSph galaxies; Figure 6 does not
rule this out, as 
the scatter around the dSph line
is large (see Section 3.2).

\vskip 5pt

\noindent{\bf 3.4 Radial Distributions}

\noindent
Figure 8 shows how the galaxies described in
the previous section ($20.5 < R < 22.5$) 
are spatially distributed in the cluster. 
Brightness encodes the number of galaxies (cluster plus
background) in the above
magnitude range in a 170 arcsecond
aperture.  The aperture size is chosen as a compromise.
Very small apertures would result in very poor counting statistics.
Very large apertures would result in such bad edge effects 
that we could not present a figure like Figure 8 over a significant
area of the cluster.
The values in neigbouring pixels are not independent because
of the aperture size; they are only truly independent in regions
more than 340 arcseconds apart; the degree of coupling between
the values falls with the distance between the pixels.   

In Figure 9 (upper panel), 
we plot the radial average of the numbers in
Figure 8, where the radius is measured from 
the center of NGC 4874 (the X-ray emission is centered on
this galaxy).
The large aperture smooths 
features in the counts distribution that are smaller than the
aperture size (73 kpc at the distance of Coma).  
Consequently, 
as outlined in the previous paragraph, adjacent 
points here are not
statistically independent; neither are the error bars, which
are in part due to counting statistics. 
In the lower panel of Figure 9, we show the corresponding
distribution for the brighter galaxies, measured in the same 
way.
The background contribution is negligible in the lower panel.

Figure 9 shows that the projected distribution of galaxies
with $20.5 < R < 22.5$ is high at the center and then
drops rapidly towards the background value between 300 
and 500 arcseconds (130 and 220 kpc).  
The true drop in the distribution may be 
sharper still, because of the smoothing due to the large
aperture size.
The radial distribution of the brighter galaxies (lower panel)
shows a much more shallow drop (density $\sim r^{-1}$;
similar results have been obtained by other
authors $-$ see e.g.~Thompson \& Gregory 1993).
Figure 8 shows that the form of the distribution for the
fainter galaxies 
is caused by two condensations of galaxies, each about 300 kpc
from NGC 4874.

The steep drop in the upper panel of Figure 9 and the
condensations seen in Figure 8 are interesting, but we stress
that this is not a big effect:
the excess galaxies seen in this central 300 kpc is approximately
30\% above the background.   
The background easily fluctuates by this amount on the
scales shown in Figure 8.
If the galaxies in the condensations in Figure 9
are cluster galaxies, 
this would suggest (i) that they have orbits which are 
more elongated towards the observer than the orbits for the giant
galaxies, or (ii) that as a population, they have a smaller velocity 
dispersion than do the larger galaxies, and so are more confined to 
the core.  If either of these is an explanation for what we
see, then this provides a very important constraint on the
formation of the Coma cluster. 
Specifically it would mean that the time since the cluster formed
(say, from a merger of two subclusters, as discussed in the
previous section) is short compared to the cluster crossing time,
or else the cluster would have virialized and
such a structure would not have survived.
In Section 3.2, we suggested obtaining velocities for the
faintest dwarfs we see there.  The same measurements would
tell us whether the condensations that we see in Figure 8 are
due to cluster or background galaxies, 
and would be able to distinguish between possibilities (i) and
(ii) above, if they were cluster galaxies.  

\vskip 5pt

\noindent{\bf 3.5 Summary and comparison with previous work}

\noindent
We have obtained a reasonably complete description of
the dwarf galaxy population in Coma down to $R\sim 24$.  A number
of important questions remain, but a velocity survey of the
central region of the cluster down to $R=24$ would be
able to address almost all of these.  Such a program will shortly
be possible with the advent of multi-object spectrographs on
8 m and 10 m telescopes.

We find a dwarf galaxy luminosity function that is steeply
rising ($\alpha \sim -1.7$) 
at faint magnitudes.  At brighter magnitudes, the
LF is flatter, with a shape that agrees well with most
previous determinations.  
In Table 7 we list a number of previous surveys
and present the LFs from these surveys in
Figure 10.
A survey which was only complete to
$M_B = -18$ or brighter would suggest a value of
$\alpha \sim -1$ at the faint-end (e.g.~Godwin \& Peach 1977,
Lugger 1986) and one complete to $M_B \sim -15$ would
suggest $\alpha \sim -1.4$ (Thompson \& Gregory 1993).  
By going deeper, we
convincingly see that the LF steepens.  In the field studied
by Bernstein et al.~(1995), this effect was masked by
an even steeper contribution from globular clusters in the
halo of NGC 4874 at the faintest magnitudes.  This is
less of a problem for us (i) because our survey covers a larger
area, and (ii) because it
excludes the halos of galaxies like NGC 4874
which have substantial globular cluster populations.
If such galaxies also have particularly high dwarf galaxy 
satellite populations, we miss those too, but careful inspection
of our data does not suggest that this is the case.  
The slope of our LF for $-14.5 < M_R < -12.5$  
is slightly steeper than that ($\alpha = -1.41$) 
measured by SH96.  Their measurement 
is not significantly different from our best
value of $\alpha$ in this range, given the (mostly systematic,
see Section 3.1 and Table 5) uncertainties.
As stated in the introduction, SH96 made completeness
corrections in their faintest points (although they were only
20\% at most before background subtraction). 
Most of our simulations suggest that these corrections
can be very unstable; for example, they depend critically on the
sizes of galaxies being corrected for. 
This might lead to systematic
uncertainties in the three faintest points of SH96 
that are larger than the random errors that they estimate. 
Then it might be possible to reconcile their data with a 
slightly steeper $\alpha$.  In any case, this difference is not
huge, and it is encouraging that their LF is so similar to ours
everywhere except at their three faintest points. 
Finally, Figure 10 suggests that the slope of $\alpha = -1.8$ for
$-19 < M_V < -15$ found by Lobo et al.~(1996) is somewhat steeper
than the slope seen in the other surveys.

Our faintest point in the $R$-band is very low ($M_R \sim -11$).  
This might
be a normalization effect, or it might be suggestive of
a turnover in the LF (most theories suggest a turnover at
faint magnitudes because of the suppression of star formation
in low mass galaxies; Efstathiou 1992, Chiba \& Nath 1994,
Thoul \& Weinberg 1995).    

The dwarf galaxies in Coma have 
colours and scalelengths that are characteristic
of dSph galaxies.  
The scale-length measurements extend the results of
Karachentsev et al.~(1995) to fainter magnitudes.
The colour distribution of galaxies with
$-14.3 < M_R < -12.3$ is strongly peaked
at $B-R = 1.3$, suggesting some homogeneity in their recent
star-formation histories. 
This is slightly bluer than what S96 found for the brighter
dwarfs; the colour gradient for the fainter galaxies is also
weaker.

The faintest dwarfs also 
seem to be more confined to the cluster center
than the giant galaxies.
Alternatively, this measurement
may in part or whole
be due to background clustering; velocity measurements
should easily distinguish between these possibilities.

We cannot make any detailed conclusions about the dwarf galaxy
population in the NGC 4839 group because the galaxy density is
too low there relative to the background.

\vskip 10pt

\noindent{\bf ACKNOWLEDGMENTS}

\noindent
This research has made use of the NASA/IPAC extragalactic database (NED) which
is operated by the Jet Propulsion Laboratory, Caltech, under agreement with the
National Aeronautics and Space Administration.

\vskip 10pt

\ni{\bf REFERENCES }
\beginrefs

Abell G.~O., 1958, ApJS, 3, 211

Babul A., Ferguson H.~C., 1996, ApJ, 458 100

Bernstein G.~M., Nichol R.~C., Tyson J.~A., Ulmer M.~P., Wittman D., 1995, AJ,
110, 1507
 
Binggeli B., 1994, in Meylan G., Prugneil P., ed., ESO
Conference and Workshop Proceedings No.~49: Dwarf Galaxies. 
European Space Observatory, Munich, p.~13

Biviano A., Durret F., Gerbal D., Le Fevre O., Lobo C., Mazure A., 
Slezak E., 1995, A\&A, 297, 610 

Briel U.~G., Henry J.~P., Bohringer H., 1992, A\&A, 259, L31
 
Burstein D., Heiles C., 1982, AJ, 87, 1165
  
Caldwell N., 1983, AJ, 88, 804
  
Cardelli J.~A., Clayton G.~C., Mathis J.~S., ApJ, 345, 245
 
Chiba M., Nath B.~B., 1994, ApJ, 436, 618
 
Coleman G.~D., Wu C-C., Weedman D.~W., 1980, ApJS, 43, 393
  
Davies J.~I., Phillipps S., Disney M.~J., 1988, MNRAS, 231, 69p

Dressler A., 1980, ApJ, 236, 351

Driver S.~P., Phillipps S., Davies J.~I., Morgan I.,
Disney M.~J, 1994a, MNRAS, 266, 155

Driver S.~P., Phillipps S., Davies J.~I., Morgan I.,
Disney M.~J, 1994b, MNRAS, 268, 393  

Efstathiou G., 1992, MNRAS, 256, 43p

Godwin J.~G., Metcalfe N., Peach J.~V., 1983, MNRAS, 202, 113

Godwin J.~G., Peach J.~V., 1977, MNRAS, 181, 323

Hodge P.~W., 1989, ARAA, 27, 139 

Hodge P., 1994, in Munoz-Tunon C., Sanchez F.,
ed., The Formation and Evolution of Galaxies.  Cambridge University
Press, Cambridge, p.~1

Hughes J.~P., Gorenstein F., Fabricant D., 1988,
ApJ, 329, 82

Jarvis J.~F., Tyson J.~A., 1981, AJ, 86, 476 
 
Jones L.~R., Fong R., Shanks T., Ellis R.~S., Peterson B.~A., 1991,
MNRAS, 249, 481.

Karachentsev I.~D., Karachentseva V.~E., Richter G.~M., Vennik J.~A.,
1995, A\&A, 296, 643
 
Kormendy J., 1985, ApJ, 295, 73

Kormendy J., 1987, in Faber S.~M. ed., Nearly Normal Galaxies. 
Springer-Verlag, New York, p.~163
 
Landolt A.~U., 1992, AJ, 104, 340

Lobo C., Biviano A., Durret F., Gerbal D., Le Fevre O., 
Mazure A., Slezak E., 1996, preprint astro-ph/9605194

Lugger P., 1986, ApJ, 303, 535

Rood H.~J., Baum W.~A., 1967, AJ, 72, 398  

Sandage A., Binggeli B., Tammann G.~A., 1985, AJ, 90, 1759

Schechter P., 1976, ApJ, 203, 297

Schombert J.~M., 1988, ApJ, 328, 475

Secker J., 1996, ApJ, 469, L81 (S96) 

Secker J., Harris W.~E., 1996, ApJ, 469, 628 (SH96)

Stewart G.~C., Fabian A.~C., Jones C., Forman W., 1984, 

Soltan A., Henry J.~P., 1983, ApJ, 271, 442

Thompson L.~A., \& Gregory S.~A., 1993, AJ, 106, 2197

Thoul A.~A., Weinberg D.~H., 1995, ApJ, 442, 480

Trentham N., 1997a, MNRAS, 286, 133 

Trentham N., 1997b, MNRAS, submitted 

Trentham N., 1997c, MNRAS, submitted 

Valdes F., 1982, Proc.~SPIE, 331, 465 

Valdes F., 1989, in Grosbol P.~J., Murtagh F., Warmels R.~H., ed.,
ESO Conference and Workshop Proceedings No.~31:
Proceedings of the 1st ESO/St-ECF Data Analysis Workshop.
European Space Observatory, Munich,  p.~35

van den Bergh S., 1976, ApJ, 206, 883  

White S.~D.~M, Kauffmann G., in Munoz-Tunon C., Sanchez F., eds.,
The Formation and Evolution of Galaxies.~Cambridge University Press,
Cambridge, p.~455

Zabludoff A. I., Huchra J.P., Geller M. J., 1990, ApJS, 74, 1

\endrefs

\vskip 10pt
 
\ni {\bf FIGURE CAPTIONS}
\vskip 10pt
\ni {\bf Figure 1.~} The Palomar Sky Survey image of the Coma cluster.
The whole image is 80 arcminutes square, with North up and East to
the left.  The two bright central galaxies are NGC 4874 (to the west)
and NGC 4889.  
The next two brightest galaxies are the anaemic (van den Bergh
1976) spiral galaxy NGC 4921, 18 arcminutes to the southeast of
NGC 4889, and NGC 4839, 41 arcminutes to the southwest of
NGC 4874.
  
The boxes represent our survey areas. 
The large box to the northeast is the Coma 
core region; it covers
674 square arcminutes, and we surveyed 97\% of it.  The smaller
box to the southwest is the NGC 4839 field. 

\vskip 10pt
\ni {\bf Figure 2.~}Number counts versus magnitude for the
Coma core region in the $R$-band (upper panel) and $B$-band
(lower panel).  The dashed lines are the mean background
counts from T97 ($\log_{10} N = 0.376 R - 4.696$ and
$\log_{10} N = 0.488 B - 7.702$).
These lines come from a fit to the number counts in T97 for
$21 < m < 25$ in both $B$ and $R$.
For $m<21$ the mean background counts shown in Figure 6 of
T97 are consistent with these lines but the errors in
the measurements there are large; any systematic
deviation of the mean counts from these lines is negligible
compared with the field-to-field variance of the background.
Brighter than $B=17$ and $R=16$, background contamination
in the Coma fields is negligible.

\vskip 10pt
\ni {\bf Figure 3.~}As Figure 2, but for the 
NGC 4839 field. 

\vskip 10pt
\ni {\bf Figure 4.~}The $R$-band luminosity function of the
Coma core region outlined in Figure 1, 
computed from the data in Figure 2 as
described in the text.  The four panels are:
(a) the LF for the whole field; (b) 
the LF for the inner 200 kpc, adopting the center of
NGC 4874 as the cluster center; (c) the LF of
the remaining part of the field not included in (b);
(d) the LF for the extreme-low-surface-brightness galaxies,
defined as galaxies whose surface brightnesses are lower than
that of any field galaxies having the same total magnitude. 

Slopes corresponding to 
different values of $\alpha$ are shown in panel (a).
The error bars represent the quadrature sum of uncertainties
from counting statistics, measurements of the total
magnitudes, and uncertainty from the field-to-field variance
of the background. 

\vskip 10pt

\ni {\bf Figure 5.~}As Figure 4, but for the $B$-band 
\vskip 10pt

\ni {\bf Figure 6.~}The 
isophotal magnitude $m_I$ and first-moment light radius $r_1$ of
all detected objects in the Coma core field.
The solid line shows where typical local dSphs placed at the
distance of Coma would lie; the scatter around this line is, however,
large.  The dashed line shows the upper envelope of the region
occupied by background galaxies (see T97).
The line is higher relative to the dSph line in the $B$-band than
in the $R$-band because the lowest-surface-brightness 
field galaxies are blue.
The dotted-dashed lines represent the line below which we detect
all zero-ellipticity exponential galaxies in the absence of other
brighter galaxies.  The lines represent the median such
line for all the images in our survey; the differences are small except
at the very faint end.  A  few objects above the line are
seen' most are either very elongated low surface-brightness galaxies,
or galaxies that we happen to detect despite our not being
complete in this region of the diagram.
\vskip 10pt

\ni {\bf Figure 7.~}Colour
distributions of the galaxies with $20.5 < R < 22.5$
($-14.3 < M_R < -12.3$ if the galaxies are in Coma)
computed as described in Section 2.2.2 using magnitudes in
a 3\sd0 diameter aperture.
When deciding whether a galaxy is included, we see if the
aperture magnitude satisfies the above criterion (except for
in panel (f), were we use the isophotal magnitude, as this
is a better approximation to the total magnitude in this case).
Only objects labeled ``galaxy'' in both $B$ and $R\,$ FOCAS
catalogs were included; this makes stellar contamination
negligible.  Cluster and background were treated the same way
throughout the analysis.
The background data come from T97.

The panels are: (a), raw data for the entire core field,
prior to background subtraction; (b),
histogram for the entire 
core field with a background contribution
subtracted; (c), as (b) but only for
galaxies with $r < 200$ kpc; (d), as (b) but for  
galaxies with $r > 200$ kpc; (e), the background-subtraccted
histogram for the NGC 4839 field, and (f) the LSB galaxies
(no background subtraction is necessary here).  

The histograms are complete to $B-R = 2.3$ (a giant elliptical in
Coma has $B-R \sim 1.9$; Coleman et al.~1980).
The vertical error bars in each bin are approximately equal to
the Poisson ($\sqrt{N}$) errors; horizontal error bars are
small.

\vskip 10pt

\ni {\bf Figure 8.~}The
number of galaxies with $20.5 < R < 22.5$
in a 170 arcsecond radius aperture
as a function of position in the Coma cluster. 
The region of the cluster core shown here is 1080 arcseconds
square; north is up and east is to the left.
The positions of NGC 4874 (X) and NGC 4889 (Y) are
shown.  The colour coding is shown below. 
\vskip 10pt

\ni {\bf Figure 9.~}The
upper panel shows the
radial average of the number of galaxies having
$20.5 < R < 22.5$
within a 170 arcsecond radius aperture, as a function of
distance from the cluster center, which we define as the
center of NGC 4874.   
The lower panel shows the corresponding plot for the
brighter galaxies ($15 < R < 19$).  The mean backgrounds
are 59.1 (the dotted line) in the upper panel and
1.13 in the lower panel.
One arcminute corresponds to 26 kpc in Coma.
\vskip 10pt

\ni {\bf Figure 10.~}The luminosity functions for the surveys
listed in Table 7.  The normalizations are appropriate to eacih
survey and are only equal for two surveys covering the same area.
Surveys covering larger area generally have lower
normalizations because they contain a bigger contribution
from the outer regions of the cluster, where the
galaxy density is low.

\par\vfill\eject\bye

%% file: stbasic.tex
\message{STBASIC.TEX TeX Macro Library}
\message{ }





\def\beginrefs{\begingroup\parindent=0pt\frenchspacing
   \parskip=1pt plus 1pt minus 1pt\interlinepenalty=1000\pretolerance=10000
   \hyphenpenalty=10000\everypar={\hangindent=0.42in}       
  \def\aa##1{{\it Astr.~Ap., \bf ##1}}
  \def\aasup##1{{\it Astr.~Ap.~Suppl., \bf ##1}}
  \def\aasupp##1{{\it Astr.~Ap.~Suppl., \bf ##1}}
  \def\aj##1{{\it A.~J., \bf ##1}}
  \def\annrev##1{{\it Ann.~Rev.\ Astr.~Ap., \bf ##1}}     
  \def\araa##1{{\it Ann.~Rev.\ Astr.~Ap., \bf ##1}}     
  \def\apj##1{{\it Ap.~J., \bf ##1}}     
  \def\apjl##1{{\it Ap.~J. (Letters), \bf ##1}}
  \def\apjlett##1{{\it Ap.~J. (Letters), \bf ##1}}
  \def\apjlet##1{{\it Ap.~J. (Letters), \bf ##1}}
  \def\apjsup##1{{\it Ap.~J.~Suppl., \bf ##1}}
  \def\apjsupp##1{{\it Ap.~J.~Suppl., \bf ##1}}
  \def\baas##1{{\it Bull.~A.A.S., \bf ##1}}
  \def\ban##1{{\it B.A.N., \bf ##1}}
  \def\ibvs##1{{\it Inf. Bull. Var. Stars}, No.~##1}
  \def\mn##1{{\it M.N.R.A.S., \bf ##1}}
  \def\mnras##1{{\it M.N.R.A.S., \bf ##1}}
  \def\pasp##1{{\it Pub.~A.S.P., \bf ##1}}
  \def\ajpasp##1{{\it Pub.~A.S.P., \bf ##1}}
  \def\nat##1{{\it Nature, \bf ##1}}
  \def\nature##1{{\it Nature, \bf ##1}}}

\def\endrefs{\endgroup}



\def\df{\leaders\hbox to 0.6em{\hss.}\hfill}


\def\section#1{\bigbreak\medskip\centerline{#1}\par\nobreak\medskip\markpage}

\def\subsection#1#2{\bigbreak\noindent{\bf#1\hskip 0.9em\relax#2}\par
   \nobreak\medskip\markpage}

\def\subsubsection#1#2{\medbreak\noindent{\sl#1\hskip 0.60em\relax#2}\par
   \nobreak\medskip\markpage}

\def\today{\advance\year by -1900 
   \number\month/\number\day/\number\year}
\def\yearmonthday{\number\year\space
   \ifcase\month\or January\or February\or March\or April\or May\or June\or
   July\or August\or September\or October\or November\or December\fi
   \space\number\day}

\newcount\num

\def\nextnum{\global\advance \num by 1 \number\num}
\def\nextitem{\leavevmode
   \hbox{\ifnum\num>8 \kern-0.43em\fi \nextnum.\kern0.60em}}
\def\bfnextitem{\leavevmode
   \hbox{\ifnum\num>8 \kern-0.43em\fi \bf\nextnum.\kern0.60em}}

\newcount\colnum

\def\nextcolnum{\global\advance \colnum by 1 \number\colnum}
\def\nextcolumn{\leavevmode
   \hbox{{\it \ifnum\colnum<9 \phantom{1}\fi Column \nextcolnum:}\kern0.60em}}

\newcount\fig

\def\nextfig{\global\advance \fig by 1 \number\fig}

\newcount\cap

\def\nextcap{\global\advance \cap by 1 \number\cap}

\newcount\letter

\def\nextlet{\global\advance \letter by 1
   \ifcase\letter\or A\or B\or C\or D\or E\or F\or G\or H\or I\or
   J\or K\or L\or M\or N\or O\or P\or Q\or R\or S\or T\or U\or V\or W\or X\or
   Y\or Z\fi}

\newdimen\bigindent \bigindent=3.5in
\def\letterhead{\hsize=6in\interlinepenalty=2000\parskip=6pt minus 3pt
  \pretolerance=750
  \def\topline##1{\hbox to\hsize{\hfil##1\hskip\rightskip}}
  \footline={\ifnum\pageno=1
    \hss\hbox{\vrule height 0.4in width 0pt}
    \eightrm Operated by the Association of Universities for Research in 
    Astronomy, Inc., for the National Aeronautics and Space Administration\hss
    \else\hfil\fi}
  \null
  \vskip-0.2in
  {\advance\rightskip by -0.75in
    \topline{3700 San Martin Drive}
    \topline{Baltimore, MD 21218}
    \topline{(301) 338-4718}\par}
  \vskip30pt minus 15pt
  {\leftskip=\bigindent\yearmonthday\par}}

\def\arpanetletterhead{\hsize=6in\interlinepenalty=2000\parskip=6pt minus 3pt
  \pretolerance=750
  \def\topline##1{\hbox to\hsize{\hfil##1\hskip\rightskip}}
  \footline={\ifnum\pageno=1
    \hss\hbox{\vrule height 0.4in width 0pt}
    \eightrm Operated by the Association of Universities for Research in 
    Astronomy, Inc., for the National Aeronautics and Space Administration\hss
    \else\hfil\fi}
  \null
  \vskip-0.2in\vskip-3\baselineskip
  {\advance\rightskip by -0.75in
    \topline{3700 San Martin Drive}
    \topline{Baltimore, MD 21218}
    \topline{(301) 338-4718}
    \topline{{\elevenrm BITNET:} \tt golombek@stsci}
    \topline{\elevenrm SPAN: \tt SCIVAX::GOLOMBEK}
    \topline{{\elevenrm ARPANET:} \tt golombek@scivax.arpa}\par}
  \vskip30pt minus 15pt
  {\leftskip=\bigindent\yearmonthday\par}}

\def\gosbletterhead{\hsize=6in\interlinepenalty=2000\parskip=6pt minus 3pt
  \pretolerance=750
  \def\topline##1{\hbox to\hsize{\hfil##1\hskip\rightskip}}
  \footline={\ifnum\pageno=1
    \hss\hbox{\vrule height 0.4in width 0pt}
    \eightrm Operated by the Association of Universities for Research in 
    Astronomy, Inc., for the National Aeronautics and Space Administration\hss
    \else\hfil\fi}
  \null
  \vskip-0.375in
  {\advance\rightskip by -0.75in
    \topline{General Observer Support Branch}
    \topline{3700 San Martin Drive}
    \topline{Baltimore, MD 21218}
    \topline{(301) 338-4996}\par}
  \vskip30pt minus 15pt
  {\leftskip=\bigindent\yearmonthday\par}}



\def\indentleft{\advance\leftskip by 50pt\interlinepenalty=750}
\def\inndentleft{\advance\leftskip by 78pt\interlinepenalty=750}
\def\narrower{\advance\leftskip by 0.42in\advance\rightskip by 0.42in
  \interlinepenalty=750}
\def\nnarrower{\advance\leftskip by 50pt\advance\rightskip by 45pt
  \interlinepenalty=750}

\def\checkbox{\nnarrower\parindent=0pt\itemitem{\vbox{\hrule height.7pt
  \hbox{\vrule width.7pt height6pt \kern6pt \vrule width.7pt}
  \hrule height.7pt}$\,$}}  


%
%
\newcount\index \index=100
\def\markpage{\advance\index by 1 \count\index=\pageno}
\def\begintableofcontents{\begingroup
  \index=100 \frenchspacing\interlinepenalty=750
  \parskip=0.1pt plus 1pt minus 0.1pt \parindent=0.3in
  \def\dfi{\advance\index by 1 \df\number\count\index}
  \def\in{\par\hskip-0.2in\indent \hangindent2\parindent \textindent}    
  \def\inin{\par\hskip0.32in\indent \hangindent3\parindent \textindent}
  \def\ininin{\par\hskip0.95in\indent \hangindent4\parindent \textindent}}



{\obeylines\gdef\startdisplay#1
  {\catcode`\^^M=5$$#1\halign\bgroup\indent##\hfil&&\qquad##\hfil\cr}}
\outer\def\enddisplay{\crcr\egroup$$}

\chardef\other=12

{\obeyspaces\gdef {\ }} 

  \font\twentyfourrm=cmr10 scaled 2488
  \font\twentyfouri=cmmi10 scaled 2074   
  \font\twentyfoursy=cmsy10 scaled 2074
  \font\twentyrm=cmr10 scaled 2074      
  \font\twentyi=cmmi10 scaled 2074   
  \font\twentysy=cmsy10 scaled 2074
  \font\eighteenrm=cmr10 scaled 1728
  \font\eighteeni=cmmi10 scaled 1728 \font\eighteensy=cmsy10 scaled 1728
  \font\fourteenrm=cmr10 scaled 1440
  \font\fourteeni=cmmi10 scaled 1440 \font\fourteensy=cmsy10 scaled 1440
  \font\twelverm=cmr12
                
  \font\twelvei=cmmi12               \font\twelvesy=cmsy10 scaled 1200
  \font\elevenrm=cmr10 scaled 1095
    
  \font\eleveni=cmmi10 scaled 1095   \font\elevensy=cmsy10 scaled 1095
  \font\tenrm=cmr10
                   
  \font\teni=cmmi10  \font\tensy=cmsy10  
  \font\ninerm=cmr9

  \font\ninei=cmmi9                  \font\ninesy=cmsy9
  \font\eightrm=cmr8
  \font\seveni=cmmi7 \font\sevensy=cmsy7

\def\commonstuff{
  \parindent=0.42in       
  \def\skipline{\vskip\baselineskip}
  \hyphenpenalty=200\pretolerance=300\tolerance=600 
  \interlinepenalty=100\clubpenalty=500\widowpenalty=500
  \nonfrenchspacing\singlespace\rm}

\def\twelvepoint{
  \font\bf=cmbx12
  \font\it=cmti12
  \font\sl=cmsl12
  \font\tb=cmtt10 scaled 1200 
  \font\tt=cmtt8 scaled 1440
  \textfont0=\twelverm \scriptfont0=\tenrm     
    \scriptscriptfont0=\sevenrm                 
  \def\rm{\fam0 \twelverm}   
  \textfont1=\twelvei  \scriptfont1=\teni  
    \scriptscriptfont1=\seveni                  
  \def\mit{\fam1 } \def\oldstyle{\fam1 \twelvei}
  \textfont2=\twelvesy \scriptfont2=\tensy 
    \scriptscriptfont2=\sevensy                 
  \def\singlespace{\baselineskip=13.5pt\lineskiplimit=-5pt
    \lineskip=0pt
    \parskip=1.25pt plus 1.5pt minus 0.25pt}  
  \def\oneandahalfspace{\baselineskip=18pt\parskip=0pt plus 1pt}
  \def\doublespace{\baselineskip=24pt\parskip=0pt plus 0.5pt}
  \footline={\ifnum\pageno=1 \hfil
             \else\hss\twelverm-- \folio\ --\hss\fi} 
  \def\pagenumbers{\footline={\hss\twelverm-- \folio\ --\hss}}  
  \def\romanpagenumbers{\footline={\hss\twelverm-- \romannumeral\folio\ --\hss}}
  \commonstuff}

\def\tenpoint{
  \font\it=cmti10
  \font\sl=cmsl10
  \font\bf=cmb10
  \textfont0=\tenrm \scriptfont0=\sevenrm     
    \scriptscriptfont0=\fiverm                 
  \def\rm{\fam0 \tenrm}   
  \textfont1=\teni  \scriptfont1=\seveni  
    \scriptscriptfont1=\fivei                  
  \def\mit{\fam1 } \def\oldstyle{\fam1 \teni}
  \textfont2=\tensy \scriptfont2=\sevensy 
    \scriptscriptfont2=\fivesy                 
  \def\singlespace{\baselineskip=12pt\lineskiplimit=0pt
    \lineskip=-0.5mm       
    \parskip=2pt plus 1pt minus 1pt}  
  \footline={\ifnum\pageno=1 \hfil
             \else\hss\tenrm-- \folio\ --\hss\fi} 
  \def\oneandahalfspace{\baselineskip=18pt\parskip=0pt plus 1pt}
  \def\doublespace{\baselineskip=24pt\parskip=0pt plus 1 pt}
  \def\pagenumbers{\footline={\hss\tenrm-- \folio\ --\hss}}  
  \def\romanpagenumbers{\footline={\hss\tenrm-- \romannumeral\folio\ --\hss}}
  \commonstuff}

\def\elevenpoint{
  \font\it=cmti10 scaled 1095
  \font\sl=cmsl10 scaled 1095
  \font\bf=cmb10 scaled 1095 
  \font\tt=cmtt10 scaled 1095
  \textfont0=\elevenrm \scriptfont0=\tenrm     
    \scriptscriptfont0=\ninerm                 
  \def\rm{\fam0 \elevenrm}   
  \textfont1=\eleveni  \scriptfont1=\teni  
    \scriptscriptfont1=\ninei                  
  \def\mit{\fam1 } \def\oldstyle{\fam1 \eleveni}
  \textfont2=\elevensy \scriptfont2=\tensy 
    \scriptscriptfont2=\ninesy                 
  \def\singlespace{\baselineskip=13pt\lineskiplimit=-5pt
    \lineskip=0mm       
    \parskip=2pt plus 1pt minus 1pt}  
  \footline={\ifnum\pageno=1 \hfil
             \else\hss\elevenrm-- \folio\ --\hss\fi} 
  \def\oneandahalfspace{\baselineskip=19pt\parskip=0pt plus 1pt}
  \def\doublespace{\baselineskip=26pt\parskip=0pt plus 1 pt}
  \def\pagenumbers{\footline={\hss\elevenrm-- \folio\ --\hss}}  
  \def\romanpagenumbers{\footline={\hss\tenrm-- \romannumeral\folio\ --\hss}}
  \commonstuff}

\def\eighteenpoint{           
  \font\bf=cmbx10 scaled 1728
  \font\it=cmti10 scaled 1728
  \font\sl=cmsl10 scaled 1728
  \font\tb=cmtt10 scaled 1728
  \font\tt=cmtt10 scaled 1728
  \textfont0=\eighteenrm \scriptfont0=\fourteenrm
    \scriptscriptfont0=\twelverm                 
  \def\rm{\fam0 \eighteenrm}   
  \textfont1=\eighteeni  \scriptfont1=\fourteeni  
    \scriptscriptfont1=\twelvei                  
  \def\mit{\fam1 } \def\oldstyle{\fam1 \eighteeni}
  \textfont2=\eighteensy \scriptfont2=\fourteensy 
    \scriptscriptfont2=\twelvesy                 
  \def\singlespace{\baselineskip=21pt\lineskiplimit=-5pt
    \lineskip=0pt
    \parskip=4pt plus 1pt minus 1pt}  
  \def\oneandahalfspace{\baselineskip=30pt\parskip=0pt plus 1pt}
  \def\doublespace{\baselineskip=40pt\parskip=0pt plus 1pt}
  \footline={\ifnum\pageno=1 \hfil
             \else\hss\eighteenrm-- \folio\ --\hss\fi} 
  \def\pagenumbers{\footline={\hss\eighteenrm-- \folio\ --\hss}}  
  \commonstuff}

\def\twentypoint{
  \font\bf=cmbx10 scaled 2074
  \font\it=cmti10 scaled 2074
  \font\sl=cmsl10 scaled 2074
  \font\tb=cmtt10 scaled 2074
  \font\tt=cmtt10 scaled 2074
  \textfont0=\twentyrm \scriptfont0=\eighteenrm     
    \scriptscriptfont0=\fourteenrm                 
  \def\rm{\fam0 \twentyrm}   
  \textfont1=\twentyi  \scriptfont1=\eighteeni  
    \scriptscriptfont1=\fourteeni                  
  \def\mit{\fam1 } \def\oldstyle{\fam1 \twentyi}
  \textfont2=\twentysy \scriptfont2=\eighteensy 
    \scriptscriptfont2=\fourteensy                 
  \def\singlespace{\baselineskip=24pt\lineskiplimit=-5pt
    \lineskip=0pt
    \parskip=5pt plus 1.5pt minus 1.5pt}  
  \def\oneandahalfspace{\baselineskip=33pt\parskip=0pt plus 1pt}
  \def\doublespace{\baselineskip=44pt\parskip=0pt plus 0.5pt}
  \footline={\ifnum\pageno=1 \hfil
             \else\hss\twentyrm-- \folio\ --\hss\fi} 
  \def\pagenumbers{\footline={\hss\twentyrm-- \folio\ --\hss}}  
  \def\romanpagenumbers{\footline={\hss\twentyrm-- \romannumeral\folio\ --\hss}}
  \commonstuff}

\def\twentyfourpoint{
  \font\bf=cmbx10 scaled 2488
  \font\it=cmti10 scaled 2488
  \font\sl=cmsl10 scaled 2488
  \font\tb=cmtt10 scaled 2488
  \font\tt=cmtt10 scaled 2488
  \textfont0=\twentyfourrm \scriptfont0=\twentyrm     
    \scriptscriptfont0=\eighteenrm                 
  \def\rm{\fam0 \twentyfourrm}   
  \textfont1=\twentyfouri  \scriptfont1=\twentyi  
    \scriptscriptfont1=\eighteeni                  
  \def\mit{\fam1 } \def\oldstyle{\fam1 \twentyfouri}
  \textfont2=\twentyfoursy \scriptfont2=\twentysy 
    \scriptscriptfont2=\eighteensy                 
  \def\singlespace{\baselineskip=28pt\lineskiplimit=-5pt
    \lineskip=0pt
    \parskip=5pt plus 1.5pt minus 1.5pt}  
  \def\oneandahalfspace{\baselineskip=42pt\parskip=0pt plus 1pt}
  \def\doublespace{\baselineskip=56pt\parskip=0pt plus 0.5pt}
  \footline={\ifnum\pageno=1 \hfil
             \else\hss\twentyfourrm-- \folio\ --\hss\fi} 
  \def\pagenumbers{\footline={\hss\twentyfourrm-- \folio\ --\hss}}  
  \def\romanpagenumbers{\footline={\hss\twentyfourrm-- \romannumeral\folio\ --\hss}}
  \commonstuff}

\def\spose#1{\hbox to 0pt{#1\hss}}
\def\lta{\mathrel{\spose{\lower 3pt\hbox{$\mathchar"218$}}
     \raise 2.0pt\hbox{$\mathchar"13C$}}}
\def\gta{\mathrel{\spose{\lower 3pt\hbox{$\mathchar"218$}}
     \raise 2.0pt\hbox{$\mathchar"13E$}}}

\def\ni{\noindent}
\def\in{\indent}
\def\inin{\in{\in}
\def\ininin{\inin{\in}}}
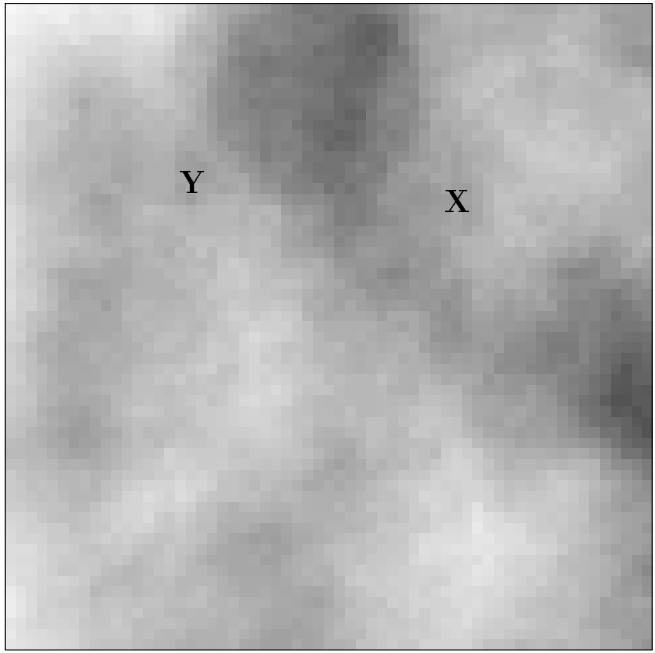
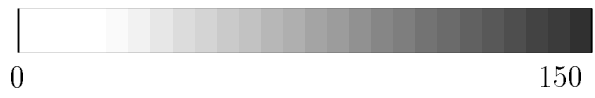
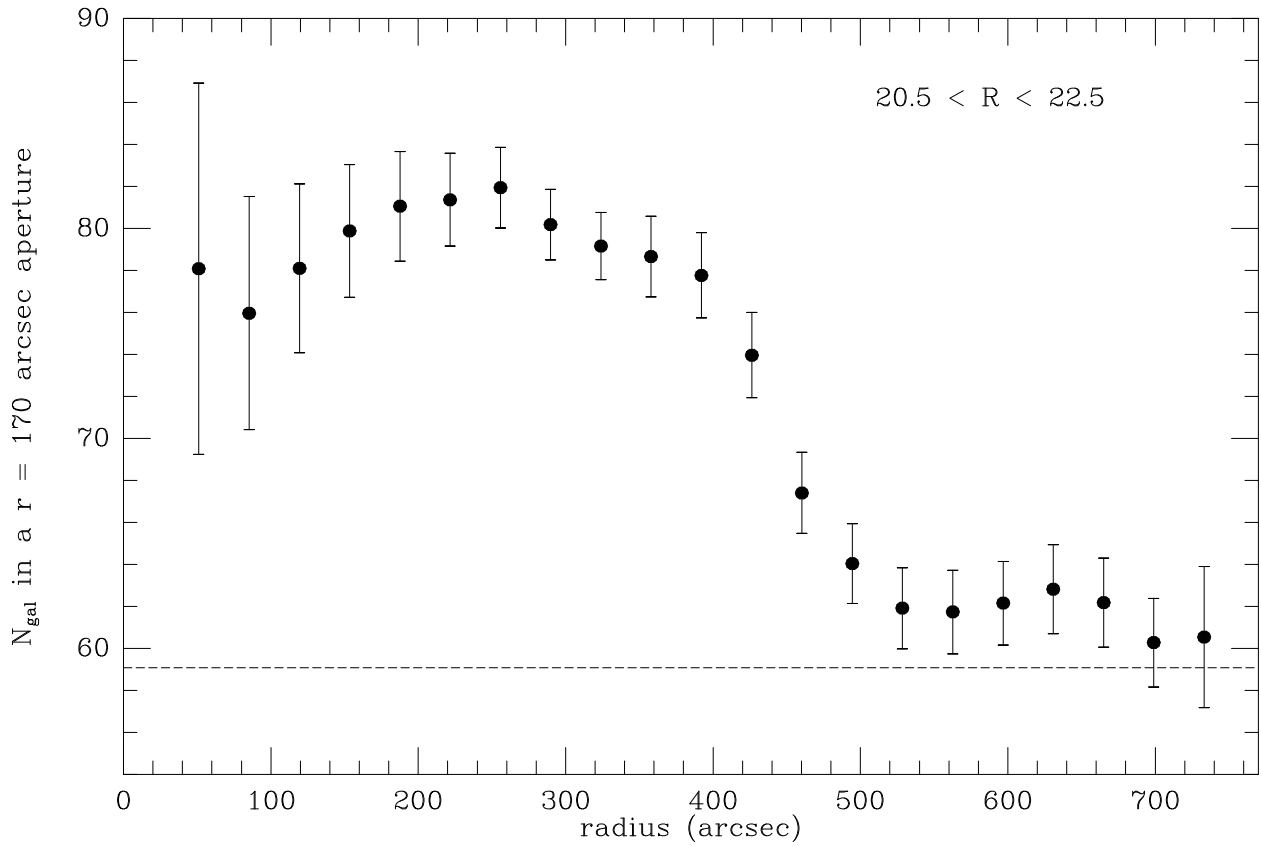
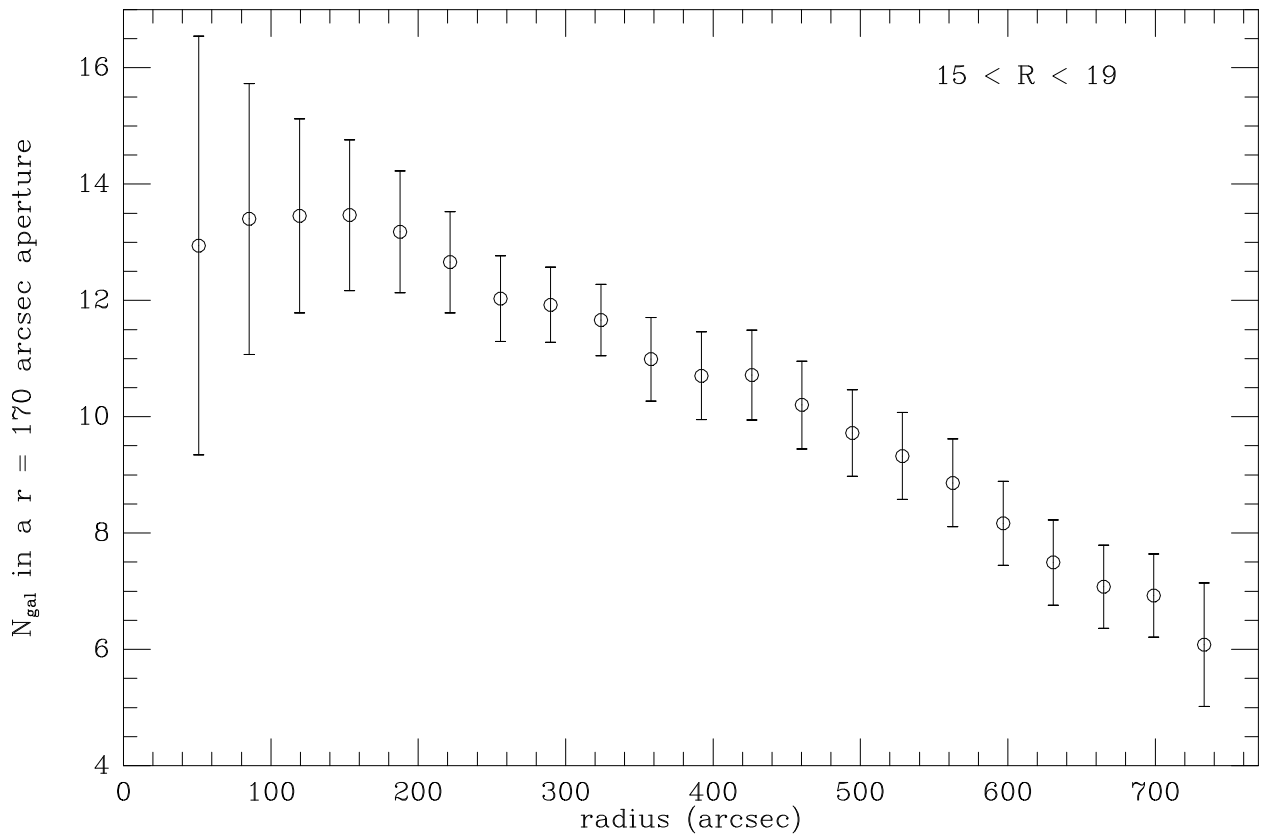
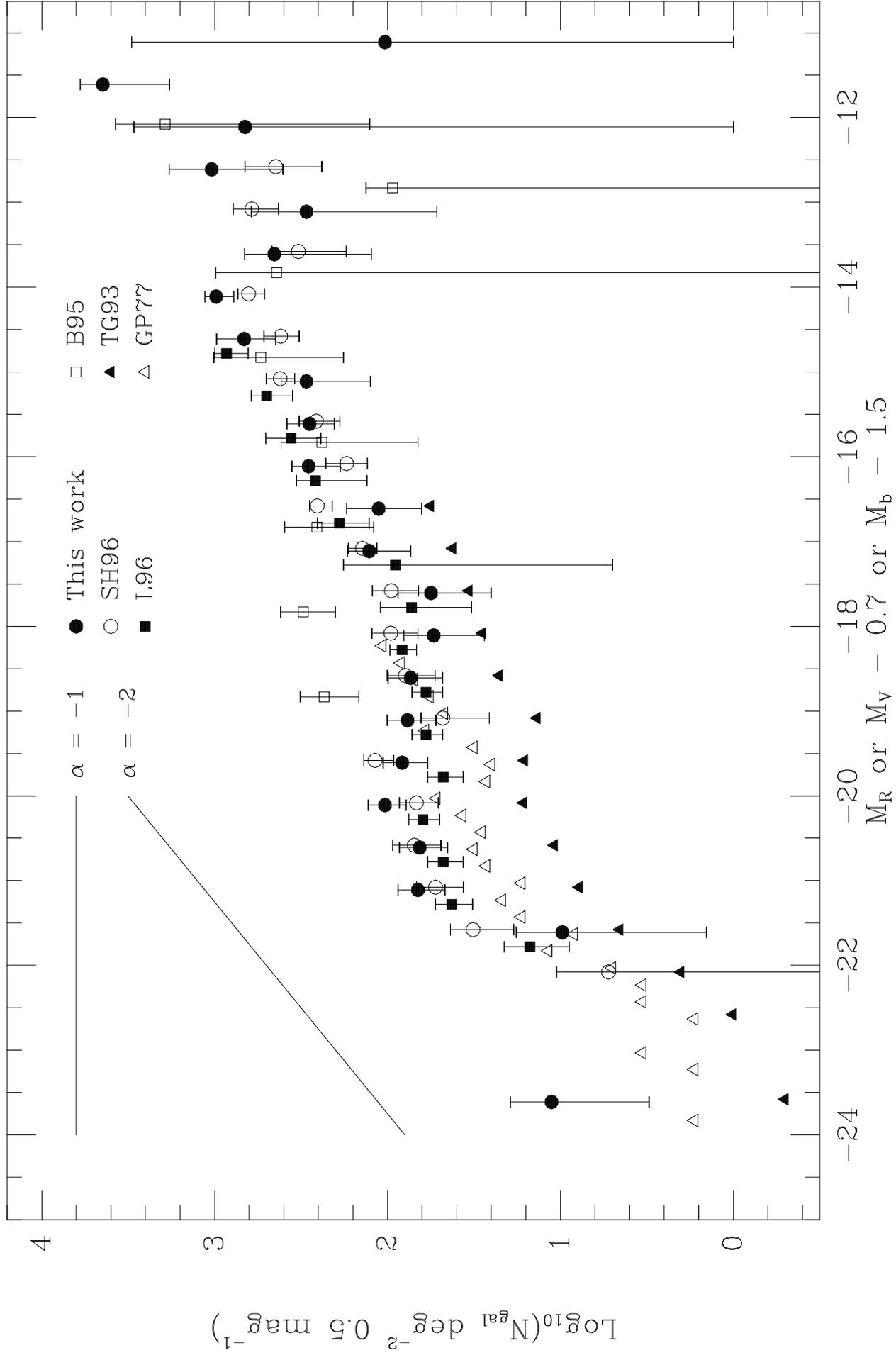
